\documentclass[runningheads]{llncs}
\usepackage[T1]{fontenc}
\usepackage{amssymb}
\usepackage{graphicx}
\usepackage{arydshln}
\usepackage{hyperref} 
\usepackage{multirow}
\usepackage{amsmath}
\usepackage{amssymb}
\usepackage{booktabs}
\usepackage{breqn}
\usepackage{array}
\newcolumntype{P}[1]{>{\centering\arraybackslash}p{#1}}
\usepackage{algorithm}
\usepackage{amsfonts}
\usepackage{graphicx,wrapfig,lipsum}
\newcolumntype{P}[1]{>{\centering\arraybackslash}p{#1}}
\usepackage[table]{xcolor}
 \usepackage{appendix}
\usepackage{caption}
\usepackage{subcaption}
\usepackage{tabularx}
\usepackage{color}

\newcommand\modelname{\texttt{BiasPruner}}
\begin{document}

\title{\modelname{}: Debiased Continual Learning for Medical Image Classification}
\titlerunning{\modelname{} for Debiased Continual Learning}
%
\author{Nourhan Bayasi\inst{1}\orcidID{0000-0003-4653-6081} \and
Jamil Fayyad\inst{2}\orcidID{0000-0003-1553-8754} \and
Alceu Bissoto\inst{3}\orcidID{0000-0003-2293-6160}\ \and
Ghassan Hamarneh\inst{4}\orcidID{0000-0001-5040-7448} \and
Rafeef Garbi\inst{1}\orcidID{0000-0001-6224-0876}}
%

\authorrunning{N. Bayasi et al.}
%
\institute{University of British Columbia, Vancouver, BC, Canada \and
University of Victoria, Victoria, BC, Canada \and 
University of Campinas, Campinas, Brazil \and
Simon Fraser University, Burnaby, BC, Canada \\
\email{nourhanb@ece.ubc.ca}}
\maketitle              
\begin{abstract} 
Continual Learning (CL) is crucial for enabling networks to dynamically adapt as they learn new tasks sequentially, accommodating new data and classes without catastrophic forgetting.  Diverging from conventional perspectives on CL, our paper introduces a new perspective wherein forgetting could actually benefit the sequential learning paradigm. Specifically, we present \modelname{}, a CL framework that intentionally forgets spurious correlations in the training data that could lead to shortcut learning. Utilizing a new bias score that measures the contribution of each unit in the network to learning spurious features, \modelname{} prunes those units with the highest bias scores to form a debiased subnetwork preserved for a given task. As \modelname{} learns a new task, it constructs a new debiased subnetwork, potentially incorporating units from previous subnetworks, which improves adaptation and performance on the new task. During inference, \modelname{} employs a simple task-agnostic approach to select the best debiased subnetwork for predictions. We conduct experiments on three medical datasets for skin lesion classification and chest X-Ray classification and demonstrate that \modelname{} consistently outperforms SOTA CL methods in terms of classification performance and fairness. Our code is available \href{https://github.com/nourhanb/BiasPruner}{here}.
\end{abstract}
\section{Introduction}
Humans inherently learn in a continual manner, acquiring new concepts over time without forgetting previous ones. In contrast, deep learning models encounter the challenge of catastrophic forgetting~\cite{lewandowsky1995catastrophic}, wherein learning new data can override previously acquired knowledge. This issue becomes especially pronounced within the medical domain, given the ever-evolving nature of medical data, the variations in acquisition protocols, the utilization of diverse devices for obtaining medical images, and other factors that contribute to shifts in data distributions or the introduction of new disease classes over time. As a result, continual learning (CL)~\cite{wang2023comprehensive,wang2023compreh} has emerged as a promising solution, allowing a network to learn continually over a sequence of presented data while forgetting as little as possible about previous knowledge. Several CL methods have emerged within the medical field to address the challenge of forgetting. Replay-based methods~\cite{perkonigg2021dynamic,kiyasseh2021clinical} store a subset of data samples and replay them to retain old information, regularization-based methods~\cite{lenga2020continual} impose restrictions on the network parameter updates to preserve prior knowledge while learning new tasks, and architecture-based methods, assign specialized architectural components for each task within the network~\cite{bayasi2023continual,bayasi2021culprit} or expand them to accommodate new tasks~\cite{gonzalez2022task}.

While previous CL methods achieved success, they have yet to consider a more realistic setting in which dataset bias exists. In medical imaging, bias could manifest through an imbalanced distribution of sensitive attributes (e.g., gender, age, ethnicity)~\cite{brown2023detecting}. Even slight imbalances induce spurious correlations between attributes and the classification target (diagnosis)~\cite{bissoto2024even}, creating an illusion of predictive power that models can exploit. Leveraging such information compromises the network’s generalization ability, amplifying societal biases over misrepresented populations in data (e.g., detecting melanoma in individuals with dark skin tones.
In CL, learning spurious correlations poses a significant challenge due to bias transfer, where biases learned by a model can be transferred to a downstream task even if it has unbiased data~\cite{salman2022does}. Since CL involves learning a sequence of tasks, the bias transfer can potentially be amplified. Moreover, recent work~\cite{busch2024truth} mathematically proved that 
handling bias becomes substantially harder when tasks are presented sequentially compared to joint training.  

\begin{figure}[ht]
    \centering
    \includegraphics[width=1\linewidth]{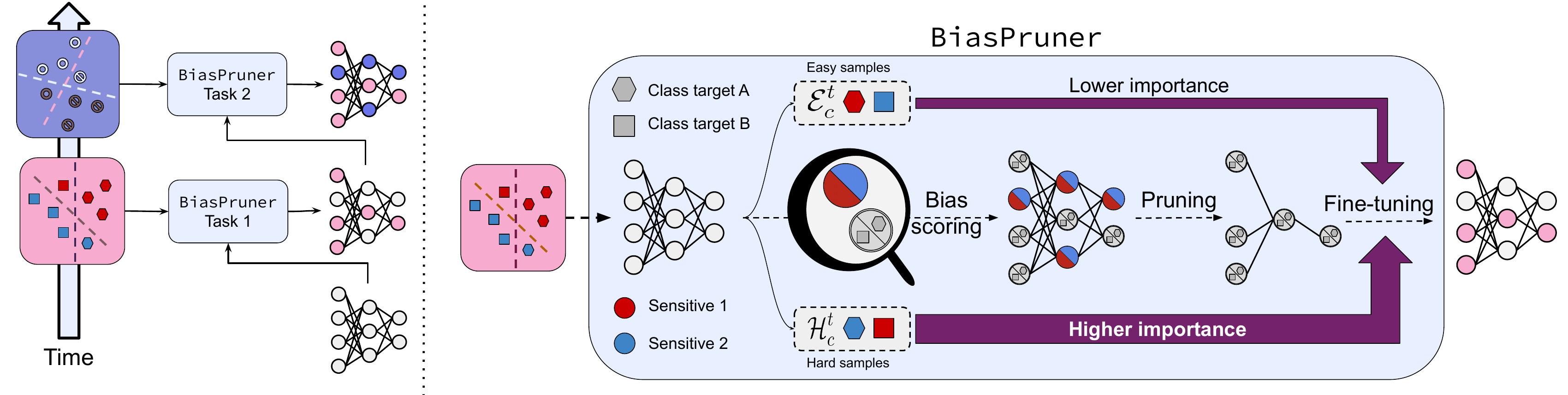}
    \caption{
    (Left) \modelname{} learns sequentially, allocating a subnetwork for each task. (Right) \modelname{} evaluates each network unit's contribution to learning spurious features from biased training data, assigning bias scores. High-score units are pruned, and the subnetwork is finetuned on both easy and hard samples.  
    }
    \label{overview}
\end{figure}

To address this gap and tackle bias in CL,  
we propose \modelname{}, a fixed-size network capable of learning sequentially and fairly over time by dedicating a unique debiased subnetwork for each task. \modelname{} leverages a newly proposed bias score to measure the contribution of each unit in the network to learning spurious features. Units with high bias scores are pruned to form a task-specific debiased subnetwork, which is kept frozen to avoid forgetting, whereas the remaining pruned units are subsequently offered for learning new tasks.  
Fig.~\ref{overview} presents an overview of our method. We evaluate our solution on three medical imaging classification datasets, each with different bias attributes. Our results demonstrate \modelname{}'s superior performance in both classification accuracy and fairness. While a few recent methods have addressed fairness in CL~\cite{lee2023continual,lesort2023spurious,chowdhury2023sustaining}, \modelname{}, to the best of our knowledge, is the first work in the medical field covering different benchmarks and bias attributes in a class-incremental setup. Crucially, \modelname{} does \textit{not} require dataset biases to be explicitly annotated. This is particularly relevant in healthcare, where identifying biases is complex and costly, compounded by patient data privacy concerns~\cite{luo2022pseudo}. 

\section{Methodology}
 \modelname{} employs a fixed-size network, $f$, capable of learning $T$ tasks sequentially, one at a time, where $T$ is not pre-determined, without forgetting any of the previously learned tasks. During training the $t$-th domain, where $t \in \{1,2,\ldots,T\}$, the network does not have access to old data, i.e., it exclusively receives biased training data $D_t = {(x_i, y_i)}$ specific to the current task, where ${(x_i, y_i)}$ represent the training samples, consisting of a total of $N_t$ images and $y_i \in \mathcal{C}_t$ classes (note the subscript $t$ emphasizing that the set of classes may change, including adding new classes, for new tasks). For clarity, we employ the symbol $c$ to denote any class within the set $\mathcal{C}_t$. \modelname{} creates a debiased subnetwork  for the $t$-th task by pruning units in the network that are mostly correlated with unknown bias(es) in $D_t$. Furthermore, \modelname{} transfers knowledge through pruning of the original network, including units of previously created  subnetworks, for each new task. At inference, \modelname{} identifies the optimal subnetwork for predictions on a given data in a task-agnostic setup; i.e., information about the task origin of a test image is unknown or unavailable.  \\ 

\noindent \textbf{2.1. Detecting Spurious Features through Bias Scoring.}
Given a biased dataset $D_t$, one of the key causes of learning shortcut predictions occurs when the model finds it easier to learn spurious features rather than the intended ones~\cite{luo2022pseudo}. Consequently, we propose to intentionally encourage the network $f$ to quickly fit on the easier features from the training data of $D_t$. To achieve this, we adopt the generalized cross entropy (GCE)~\cite{zhang2018generalized} loss, $\mathcal{L}_{\mathrm{GCE}}$, which was originally proposed to address noisy labels by fitting on the easier clean data and slowly memorizing the hard noisy samples. The GCE loss is formulated as follows: 
$$
\mathcal{L}_{\mathrm{GCE}}(p(x ; \theta), y)=\frac{1-p_y(x ; \theta)^q}{q}
$$
\noindent where $q \in(0,1]$ is a hyperparameter controlling the degree of bias amplification, $p(x ; \theta)$ and $p_y(x ; \theta)$ are the softmax output of the network and its probability assigned to the target label $y$, respectively. Due to the GCE loss's gradient, which up-weights samples with a high probability of predicting the correct target, the network quickly becomes biased to easier samples and learns shortcuts~\cite{nam2020learning}. 

Once the network is biased, it becomes logical to identify the units that have contributed the most to learning the shortcut in each class.  To achieve this, we partition the training data $\{x, y\}$ into two groups per groundtruth class $c$: The biased sample set, $\mathcal{E}_{c}^{t}$, consists of $(x_i, y_i)$ pairs that are {correctly} classified by the biased network with a probability $p_{y,i}\geq \tau$; i.e.,  samples that are $\mathcal{E}$asier for the network to learn. Similarly, the unbiased sample set, $\mathcal{H}_{c}^{t}$, comprises $(x_i, y_i)$ pairs that are  {misclassified} by the biased network; i.e., samples that are $\mathcal{H}$arder to learn, as follows  (Refer to supplementary material,  Fig.~\ref{babc}, for visualizations): 
$$
\mathcal{E}_{c}^{t} = \{i \,|\, y_i = c_i \, ~\&~ \, p_{y,i} \geq \tau\} ~~, ~~ \mathcal{H}_{c}^{t} = \{i \,|\, y_i \neq c_i \, ~\&~ \, p_{y,i} \geq \tau\} .
$$ 

Next, a we define a bias score, $\mathcal{S}^t_{c, n}$, for each unit $n$ in the biased network relative to a given class $c$ by analyzing each unit's ReLU activation, $a_i^n$, as follows:
$$
\mathcal{S}^t_{c, n} = \frac{1}{\left|\mathcal{E}_{c}^{t}\right|} \sum_{i \in \mathcal{E}_{c}^{t}} \text{Var}\left(a_i^n\right) - \frac{1}{\left|\mathcal{H}_{c}^{t}\right|} \sum_{i \in \mathcal{H}_{c}^{t}} \text{Var}\left(a_i^n\right)
.$$
\noindent  \(\text{Var}\left(a_i^n\right)\) represents the variance of the feature map \(a_i^n\) over its spatial dimensions \((w, h)\). The final unit-based bias score $\bar{\mathcal{S}^t_{n}}$  is calculated by averaging the results over all class-specific scores. Units that respond more strongly to biased samples (c.f. $\mathcal{E}_{c}^{t}$) than to unbiased samples (c.f. $\mathcal{H}_{c}^{t}$) are assigned higher bias scores, designating them as the main contributors to learning shortcuts in the network. \\

\noindent \textbf{2.2 Forming Subnetworks by Bias-aware Pruning and Finetuning.} To ensure fairness in CL, we form a task-specific, debiased subnetwork, ${f}_t$,  by selectively removing the units responsible for learning the bias in $D_t$. The pruning involves removing the top $\gamma\%$ of units, which includes the output feature maps with the highest bias scores and their corresponding filters, leaving ($1-\gamma\%$) for each \(f_t\). To counteract potential performance drop post-pruning while prioritizing improved performance on harder-to-learn samples, we propose a new weighted cross entropy loss, $\mathcal{L}_{\mathrm{W C E}}$, for fine-tuning $f_t$ on $D_t$ over a few epochs: 
$$\mathcal{L}_{\mathrm{W C E}}(x) = \mathcal{W}(x) \cdot \mathcal{L}_{\mathrm{C E}}\left({f}(x), y\right), ~~~\text{where} ~~~ \mathcal{W}(x) =\exp \left(\alpha \cdot \mathcal{L}_{\mathrm{G C E}}(x) \right) .
$$
\noindent  \(\alpha \in (0,1)\) is a trainable parameter, and $\mathcal{L}_{\mathrm{G C E}}(x)$ is the sample's GCE loss value determined as discussed in Sec. 2.1. With this weighted function,  the influence of training samples in the finetuning process varies according to their bias alignment; i.e., easy samples (c.f., $\mathcal{L}_{\mathrm{G C E}}(x)$ is small) are down-weighted, whereas hard samples  (c.f., $\mathcal{L}_{\mathrm{G C E}}(x)$ is large)  are up-weighted, exponentially. \\

\noindent \textbf{2.3. Debiased Knowledge Transfer for Enhanced Task Adaptation.} When learning a new task, \modelname{} facilitates knowledge transfer (KT), which is achieved by pruning the entire original network $f$ to create the new task-specific subnetwork, including both free units and pre-assigned debiased subnetworks of previous tasks. To avoid forgetting the previously acquired knowledge, the subnetworks associated with prior tasks are kept frozen and only the free units are updated to learn the new task. \\

\noindent \textbf{2.4. Task-agnostic Inference.} 
\modelname{} addresses a practical scenario where the task identity of a test image is unknown during inference. In other words, the specific task to which an image belongs to is not explicitly provided. Given a test batch of size $s$ as $\mathbf{X}^{\text {test }}$, we employ a `maxoutput' strategy for task prediction~\cite{dekhovich2023continual}, which involves identifying the task with the maximum output response:
$ t^*=\underset{t=1,2, \ldots, T}{\arg \max } \sum_{i=1}^s \max \varphi_t\left(\theta_t\left(\mathbf{x}_i^{t e s t}\right)\right) 
$, where $\varphi_t$ is the fully connected layer of the $t$-th subnetwork. Subsequently, we use the selected  $t^*$ task to make the final prediction $\hat{\mathbf{y}}$ based on the corresponding subnetwork; $\hat{\mathbf{y}}={f}_{t^*}\left(\mathbf{X}^{\text {test }}\right) 
$.

\section{Experiments and Results} \label{results_section}
\noindent \textbf{Datasets.} We selected/constructed datasets based on three primary considerations: (a) the presence of a dataset bias that is spuriously correlated with the disease classes; (b) The need for a variety of classes to facilitate the CL setup; and (c) publicly available to ensure reproducibility.  Hence, we include Fitzpatrick17K (FITZ)~\cite{groh2021evaluating}, HAM10000 (HAM)~\cite{ham} and NIH ChestX-Ray14 (NIH)~\cite{wang2017chestx}. Each dataset has 114, 7 and 14 distinct classes, respectively, that are split into 6, 3 and 3 tasks, respectively, with non-overlapping classes, as shown in Table~\ref{tab2} (Refer to supplementary material for dataset (Table~\ref{nih1}) $\&$ bias (Fig.~\ref{appendix_plot}) details). 

\begin{table}[t]
\centering
\scriptsize
\caption{Details on the multi-class disease datasets used in our experiments. }
\begin{tabular}{c|ccccc}
\hline 
{Dataset}& {~Number of images~} & {Classes} & {~Tasks~} & {Classes per task } & {Dataset bias} \\
\hline 
FITZ & 16,012	& 114 & 6 & {$[19, 19, 19, 19, 19, 19]$} & ~~Skin tone (I, II, III, IV, V, VI)~  \\
HAM &  8,678 &	 7 & 3 & {$[2, 2, 3]$} & Age (age$\geq$60, age$<$60) \\
NIH & 19,993&	 14 & 3 & {$[4, 5, 5]$} & Gender (male, female) \\ \hline
\end{tabular}
\label{tab2}
\end{table}

\noindent \textbf{Evaluation Metrics.} We assess the performance of \modelname{} using both the accuracy and fairness metrics. We use the commonly used F1-score (F) and balanced accuracy (ACC) metrics. We report the accuracy per sensitive attribute (e.g., male, female) as well as overall class performance (Overall). For fairness, we use the demographic parity ratio (DPR) and equal opportunity difference (EOD) metrics.  Similar to other CL methods, we report all metrics at the end of learning (i.e., after training the model on all $T$ tasks), averaged across all tasks.

\noindent \textbf{Implementation Details.}
We use ResNet-50~\cite{he2016deep}  as the backbone for feature extraction and a unified classifier for all tasks during inference. We use the Adam optimizer with a batch of 32 images for 200 epochs to train \modelname{} with $\mathcal{L}_{\mathrm{G C E}}$, having early stopping in case of overfitting. We set $q$ in $\mathcal{L}_{\mathrm{GCE}}$ to 0.7 (default) and the confidence threshold to $\tau=0.70$. We set the pruning ratio to $\gamma=0.6$ for all tasks. For the finetuning with $\mathcal{L}_{\mathrm{W C E}}$, we train the debiased subnetwork for 20 epochs, and we saved the weights with the highest ACC and EOD on the validation set. In all experiments, we report averaged results across three random task orders, aiming to neutralize any potential impact of the order in which tasks are processed during network training. 

\noindent \textbf{I. Quantitative Results on Skin-tone-biased Dataset}  (FITZ) are reported in Table~\ref{results_fitz}. First, we compare \modelname{} (Exp~$\mathcal{D}$) against three common CL baselines (Exp~$\mathcal{A}$): {JOINT}, which consolidates data from all tasks for joint model training; {SINGLE}, which trains separate models for each task and deploys task-specific models during inference; and {SeqFT}, which finetunes a single model on the current task without addressing forgetting. We observe that SINGLE outperforms JOINT as each task is learned independently, leading to improved classification and fairness results, and that SeqFT exhibits a significant performance drop due to catastrophic forgetting. Notably, \modelname{} outperforms baselines in terms of overall accuracy and fairness,  attributing this to its ability to reduce the training data bias and transfer knowledge across the tasks. 

 \begin{table}[t]
\centering
\scriptsize
\caption{Classification performance and fairness on FITZ.  Best results marked in \textbf{bold} (except upper-bound). Higher is better for all metrics except EOD.}
\begin{tabular}{c|c|cccccccccc}
\hline 
\multirow{2}{*}{Exp~}  & \multirow{2}{*}{Method}  & \multirow{2}{*}{F} &  \multicolumn{7}{c}{ACC}   & \multirow{2}{*}{DPR} & \multirow{2}{*}{EOD} \\ \cline{4-10}
 &  &   & Type-I &	Type-II&	Type-III&	Type-IV&	Type-V	&Type-VI	&Overall &		&  \\ \hline
 \multicolumn{12}{c}{\textbf{Comparison against Baselines}} \\ \hline
 \multirow{3}{*}{$\mathcal{A}$}  & JOINT & 0.256	&	0.269	&	0.304	&	0.335	&	0.309	&	0.365	&	0.245	&	0.324	&	0.137	&	0.298  \\ 
& SINGLE & 0.435	&	0.410	&	0.469	&	0.465	&	0.495	&	0.492	&	0.430	&	0.472	&	0.185	&	0.251	\\
& SeqFT & 0.188	&	0.187	&	0.261	&	0.299	&	0.254	&	0.214	&	0.192	&	0.221	&	0.051	&	0.721 \\ \hline
 \multicolumn{12}{c}{\textbf{Comparison against CL Methods}} \\ \hline
\multirow{3}{*}{$\mathcal{B}$} & EWC & 0.325	&	0.254	&	0.356	&	0.355	&	0.401	&	0.412	&	0.244	&	0.324	&	0.212	&	0.342  \\ 
 & PackNet & 0.433	&	0.366	&	0.402	&	0.445	&	0.447	&	0.479	&	0.319	&	0.414	&	0.154	&	0.425  \\ 
&  SupSup & 0.451	&	0.254	&	0.298	&	0.441	&	0.452	&	0.436	&	0.410	&	0.425	&	0.162	&	0.431 \\ \hline
\multicolumn{12}{c}{\textbf{Comparison against CL with Bias Mitigation Methods}} \\ \hline
\multirow{6}{*}{$\mathcal{C}$} & EWC+S & 0.308	&	0.264	&	0.357	&	0.324	&	0.411	&	0.417	&	0.385	&	0.341	&	0.228	&	0.311   \\ 
& PackNet+S &0.495	&	0.434	&	0.485	&	\textbf{0.494}	&	\textbf{0.565}	&	0.562	&	0.584	&	0.501	&	0.184	&	0.248  \\
&SupSup+S &0.466	&	0.418	&	0.467	&	0.432	&	0.554	&	{0.561}	&	0.534	&	0.492	&	0.182	&	0.221  \\
&EWC+W & 0.321	&	0.251	&	0.356	&	0.334	&	0.392	&	0.401	&	0.398	&	0.346	&	0.216	&	0.298  \\ 
& PackNet+W & 0.527	&	0.405	&	0.477	&	0.480	&	0.529	&	0.546	&	0.524	&	0.472	&	0.144	&	0.246   \\ 
&SupSup+W  &0.457	&	0.425	&	0.451	&	0.448	&	0.530	&	0.561	&	0.544	&	0.508	&	0.178	&	0.254   \\ \hline
\multicolumn{12}{c}{\textbf{Our Proposed Fair Continual Learning Method}} \\ \hline
$\mathcal{D}$ &\modelname{} &\textbf{ 0.540}	&	\textbf{0.457}	&	\textbf{0.502}	&	0.435	&	0.551	&	\textbf{0.563}	&	\textbf{0.584}	&	\textbf{0.512}	&	\textbf{0.331}	&	\textbf{0.202 }  \\\hline \hline
\multicolumn{12}{c}{\textbf{[Upper-bound] Comparison against a Bias Mitigation Method}} \\ \hline
$\mathcal{E}$& FairDisCo &0.542	&	0.479&		0.523		&0.468&		0.571&		0.574&		0.615&		0.548	&	0.474&		0.192    \\ \hline
\end{tabular}
\label{results_fitz}
\end{table}

Secondly, we compare \modelname{} against three CL methods (Exp~$\mathcal{B}$): EWC~\cite{kirkpatrick2017overcoming}, a regularization-based method, and PackNet~\cite{mallya2018packnet} and SupSup~\cite{wortsman2020supermasks}, both subnetwork-based like ours. 
We notice that these CL methods demonstrate lower fairness compared to \modelname{}, which is expected as they overlook dataset bias. Specifically, PackNet and SupSup exhibit higher accuracy but lower fairness compared to EWC. This is mainly due to their subnetwork-based nature, which can inadvertently worsen accuracy disparities, particularly among specific subgroups, during the removal of unimportant parameters~\cite{lin2022fairgrape}. 

Thirdly, we enhance the competing CL methods by augmenting each of them with pre-processing bias mitigation algorithms (Exp~$\mathcal{C}$). Specifically, we apply the Resampling Algorithm (S), which balances the dataset by oversampling minorities and undersampling majorities within each pair of skin label and tone. Additionally, we explore the Reweighting Algorithm (W)~\cite{du2022fairdisco}, which assigns lower weights to images that have been disadvantaged or favored to prevent the model from learning discriminatory features.  While showing improved accuracy and fairness compared to Exp~$\mathcal{B}$, they fall short of our \modelname{}'s performance.

Finally, we compare \modelname{} to FairDisCo~\cite{du2022fairdisco}, a (non-CL) bias mitigation technique for medical applications, which uses bias annotations in training. Therefore, it can set an upper bound on the performance. For a fair comparison,   we allow FairDisCo to learn each task independently and report the average performance over all tasks (Exp~$\mathcal{E}$). Despite not using bias annotations, \modelname{} exhibits slightly lower but comparable performance to FairDisCo.

\noindent \textbf{II. Quantitative Results on Age- and Gender-biased Dataset}  (HAM \& NIH, respectively) are given in Table~\ref{results_nih}.  \modelname{} (Exp~$\mathcal{I}$) outperforms other baselines (Exp~$\mathcal{F}$), CL methods (Exp~$\mathcal{G}$),  and CL methods with debiasing (Exp~$\mathcal{H}$) in both overall task classification and fairness. 

 \begin{table}[t]
\centering
\scriptsize
\caption{Classification performance and fairness on HAM and NIH. Best results marked in \textbf{bold} (except upper-bound). }
\begin{tabular}{c|c|cccccc|cccccc}
\hline
\multirow{3}{*}{\rotatebox{90}{Exp}} & \multirow{3}{*}{Method}  &  \multicolumn{6}{c|}{HAM}   & \multicolumn{6}{c}{NIH} \\ \cline{3-14}
 &  & \multirow{2}{*}{F} &  \multicolumn{3}{c}{ACC}   & \multirow{2}{*}{DPR} &  \multirow{2}{*}{EOD} & \multirow{2}{*}{F} &  \multicolumn{3}{c}{ACC}   & \multirow{2}{*}{DPR} &  \multirow{2}{*}{EOD} \\ \cline{4-6} \cline{10-12}
  &  &  & $<$60 &$\geq$60&	Overall &	&  & &  M &	F&	Overall &	&  \\ \hline
 \multicolumn{14}{c}{\textbf{Comparison against Baselines}} \\ \hline
  \multirow{3}{*}{$\mathcal{F}$} & JOINT  & 0.755&	0.781	&0.665	&0.738&	0.239	&0.320 	&	0.282	&	0.306	&	0.259	&	0.285	&	0.706	&	0.325  \\   
& SINGLE & 0.836&	0.819&	0.834&	0.841&	0.609&	0.131 &	0.434	&	0.428	&	0.403	&	0.417	&	0.728	&	0.311  \\ 
& SeqFT & 0.431&	0.372	&0.404&	0.416	&0.201&	0.558 &	0.219	&	0.251	&	0.217	&	0.231	&	0.246	&	0.544 \\ \hline
\multicolumn{14}{c}{\textbf{Comparison against CL Methods}} \\ \hline
\multirow{3}{*}{$\mathcal{G}$}  & EWC&	0.788	&0.773	&0.804&	0.772&	0.561&	0.360&	0.398	&	0.428	&	0.405	&	0.417	&	0.562	&	0.264 \\
& PackNet&	0.824	&0.807	&0.799&	0.808	&0.620&	0.302&	0.434	&	0.47	&	0.444	&	0.458	&	0.588	&	0.284 \\
& SupSup&	0.831 &	0.788&	0.845&	0.822&	0.625&	0.296&	0.448	&	0.451	&	0.441	&	0.445	&	0.571	&	0.293 \\ \hline
\multicolumn{14}{c}{\textbf{Comparison against CL with Bias Mitigation Methods}} \\ \hline

\multirow{6}{*}{$\mathcal{H}$}  & EWC+S &0.834	&0.821&	0.832	&0.827&	0.575&	0.172&	0.412	&	0.434	&	0.416	&	0.421	&	0.567	&	0.259   \\ 
&PackNet+S  & 0.839	&0.849&	0.817	&0.829&	0.613	&0.181 &	0.419	&	0.44	&	0.425	&	0.434	&	0.640	&	0.211  \\
&SupSup+S & 0.849&	0.802&	0.811&	0.817&	0.639&	0.204& 	0.432	&	0.456	&	0.448	&	0.451	&	0.662	&	0.204 \\ 
&EWC+W & 0.791&	0.778	&0.784	&0.781&	0.544&	0.168	&	0.418	&	0.441	&	0.423	&	0.432	&	0.569	&	0.251 \\ 
 &PackNet+W & 0.814&	\textbf{0.877}	&0.819&	0.842&	0.549&	0.189 &	0.443	&	0.462	&	0.456	&	0.459	&	0.704	&	0.192 \\ 
&SupSup+W   &0.846	&0.797&	0.809&	0.803	&0.536&	0.213	&	0.458	&	0.481	&	0.463	&	0.474	&	0.731	&	\textbf{0.184 }  \\ \hline
\multicolumn{14}{c}{\textbf{Our Proposed Fair Continual Learning Method}} \\ \hline
$\mathcal{I}$ & \modelname{}  & \textbf{0.860} &	0.851	&\textbf{0.852}&	\textbf{0.858}&\textbf{	0.642}&	\textbf{0.127}	&	\textbf{0.488}	&	\textbf{0.525}	&	\textbf{0.484}	&	\textbf{0.507}	&	\textbf{0.821}	&	{0.188 }  \\ \hline \hline
\multicolumn{14}{c}{\textbf{[Upper-bound] Comparison against a Bias Mitigation Method}} \\ \hline
$\mathcal{J}$ & FairDisCo  &0.873	&0.876&	0.904&	0.893	&0.682&	0.113 &	0.486&	0.545&	0.512&	0.538&	0.855&	0.150 \\  \hline
\end{tabular}
\label{results_nih}
\end{table}

\noindent \textbf{III. Ablation Studies} analyze the impact of individual components in \modelname{} (Table~\ref{ablation}). In Exp~$\mathcal{K}$, we train the model using CE loss instead of GCE. In Exp~$\mathcal{L}$, we randomly prune the network instead of using our bias-based pruning. In Exp~$\mathcal{M}$, we finetune the debiased subnetworks in \modelname{} with CE loss without weighting it. In Exp~$\mathcal{N}$, we simulate the absence of knowledge transfer (KT) by prohibiting any overlapping between the parameters $\theta_t$ and $\theta_{t'}$ for any two tasks $t$ and $t'$. We observe that 
1) the impact of $\mathcal{L}_{\mathrm{W C E}}$ (Exp~$\mathcal{M}$) is predominant, as fine-tuning with CE leads to the poorest performance in accuracy and fairness,  attributed to the risk of subnetworks potentially relearning bias. 

 \begin{table}[t]
\centering
\scriptsize
\caption{Classification (Overall) $\&$ fairness (DPR) results of \modelname{} from ablation studies. Best results marked in \textbf{bold}.}
\begin{tabular}{c|cccc|cc|cc|cc}
\hline
\multirow{2}{*}{\rotatebox{0}{Exp}} &  \multirow{2}{*}{~$\mathcal{L}_{\mathrm{G C E}}$}~ & ~\multirow{2}{*}{Bias-aware Pruning}~ & ~\multirow{2}{*}{$\mathcal{L}_{\mathrm{W C E}}$}~ &  ~\multirow{2}{*}{KT}~ & \multicolumn{2}{c|}{FITZ} & \multicolumn{2}{c|}{HAM} & \multicolumn{2}{c}{NIH} \\ \cline{6-11}
& & & & &  Overall$\uparrow$ & ~DPR$\uparrow$ & Overall$\uparrow$ & ~DPR$\uparrow$& Overall$\uparrow$ & ~DPR$\uparrow$ \\ \hline
$\mathcal{D,I}$&\checkmark &  \checkmark & \checkmark & \checkmark& \textbf{0.512}	&\textbf{0.331}	&\textbf{0.858}	&\textbf{0.642}	&\textbf{0.507}	&\textbf{0.821} \\ \hline
$\mathcal{K}$&$\times$ & \checkmark & \checkmark &\checkmark &0.498&	0.254&	0.834&	0.579&	0.501&	0.779 \\ 
$\mathcal{L}$&\checkmark & $\times$ &  \checkmark &\checkmark & 0.508	&0.328	&0.842	&0.637&	0.498&	0.814 \\ 
$\mathcal{M}$&\checkmark &  \checkmark & $\times$ & \checkmark& 0.481 &	0.247&	0.792&	0.576	&0.468	&0.754 \\ 
$\mathcal{N}$&\checkmark &  \checkmark &  \checkmark& $\times$  &0.504 &	0.324&	0.851	&0.630&	0.496&	0.803 \\ \hline
\end{tabular}
\label{ablation}
\end{table}

\begin{figure}[t]
    \centering
\includegraphics[width=0.92\linewidth]{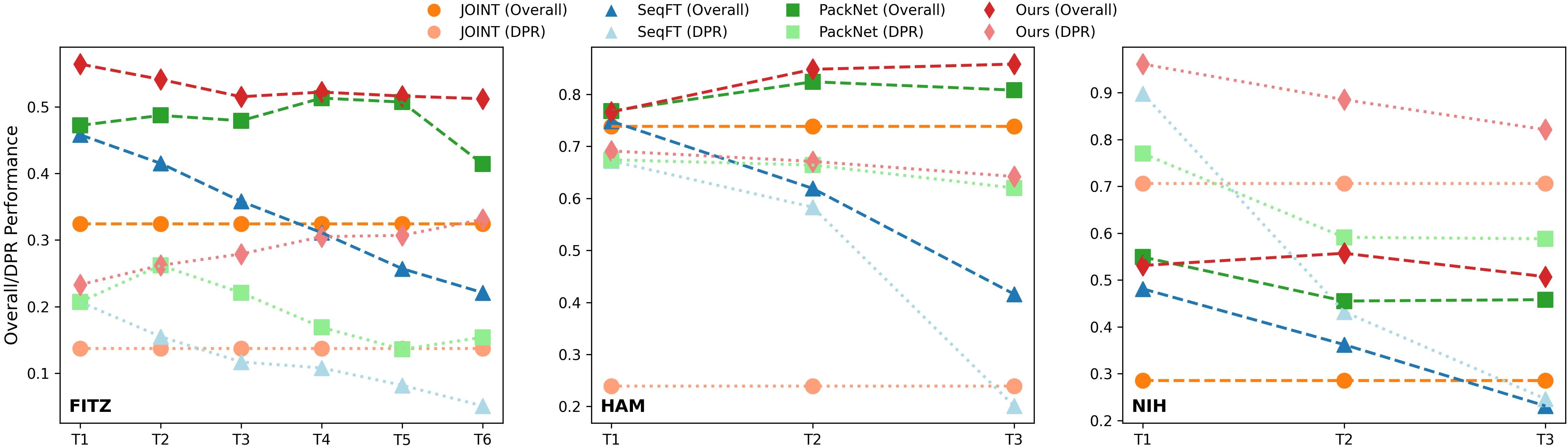}
    \caption{The overall (dashed) and DPR (dotted) performance of \modelname{} and other methods over \textit{all the
seen tasks} after each training step in the continual learning sequence, where $Ti$ refers to the $i$th task.}
    \label{clanalysis}
\end{figure}

\noindent \textbf{IV. A Sequential Analysis}, as illustrated in Fig.~\ref{clanalysis}, showcases the consistently superior performance of \modelname{} over other methods in terms of overall accuracy and DPR  after each step in the continual learning sequence across FITZ, HAM, and NIH datasets. 

\begin{figure}[H]
    \centering
\includegraphics[height=2.1cm,width=0.92\linewidth]{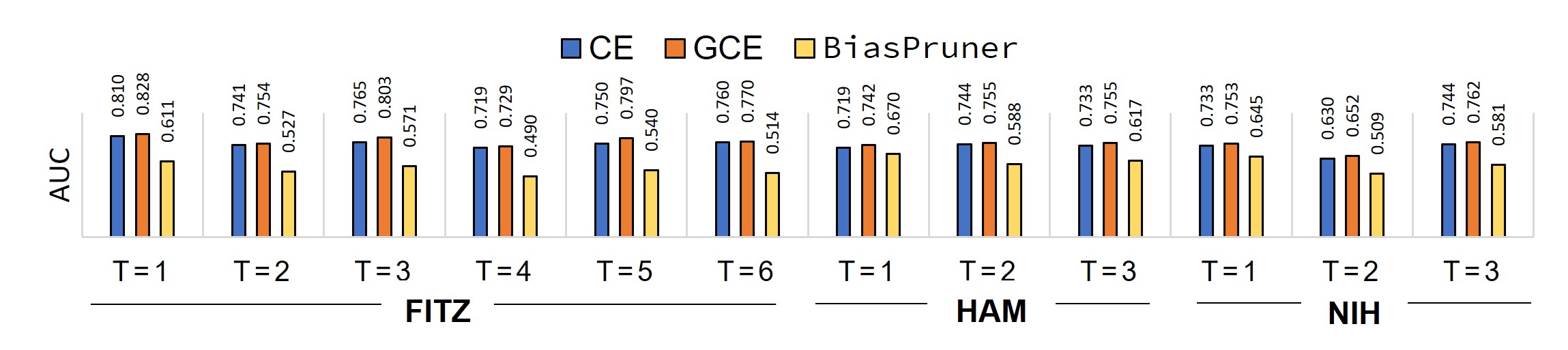}
    \caption{Sensitive attribute detection from frozen models pre-trained to diagnose. \modelname{} low AUCs indicate that bias is not encoded in its resulting features.}
    \label{freeze}
\end{figure}

\noindent \textbf{V.} For \textbf{Analysis of Model Biases} (e.g., skin tone), we trained classifiers for sensitive attribute detection on top of frozen feature extractors from three networks: CE-based, GCE-based, and our \modelname{}, all pre-trained to diagnose (Fig.~\ref{freeze}). The better-than-chance ($\in$ [0.63, 0.828]) detection accuracy of sensitive attributes, in CE- and GCE-based training, reveals that sensitive data is embedded in the originally learned features, i.e., the presence of bias. GCE, due to its loss function promoting shortcut learning, showed the most bias. The high accuracy achieved by CE shows that even naïvely trained models are susceptible to bias. In contrast, \modelname{} shows minimal bias, reflected by its near-random AUC values  ($\in$ [0.49, 0.67]) when detecting sensitive information. 

%

\section{Conclusion}
In this paper, we presented \modelname{}, a new continual learning (CL) framework that leverages intentional forgetting to improve fairness and mitigate catastrophic forgetting. By quantifying the influence of each network unit on learning spurious features, \modelname{} identifies and prunes highly biased units to construct a debiased subnetwork for each task. Experimental evaluations on three classification datasets demonstrate that \modelname{} consistently outperforms baselines and CL methods in classification performance and fairness. Our results highlight the need to address dataset bias in future CL designs and evaluations, due to the frequent fairness shortcomings of current CL methods.

\subsubsection{\ackname} We thank NVIDIA for their hardware grant and the Natural Sciences and Engineering Research Council (NSERC) of Canada for the Vanier PhD Fellowship. A. Bissoto is funded by FAPESP (2019/19619-7, 2022/ 09606-8). 

\subsubsection{\discintname}
The authors have no competing interests to declare that are relevant to the content of this article.
 \bibliographystyle{splncs04}
 \bibliography{mybibliography}
\newpage
\appendix
\begin{center}
    \noindent \large{\textbf{{\modelname{}: Debiased Continual Learning for Medical Image Classification (Supplementary Material)}}}
\end{center}
\begin{figure}[H]
    \centering
    \includegraphics[width=1\linewidth]{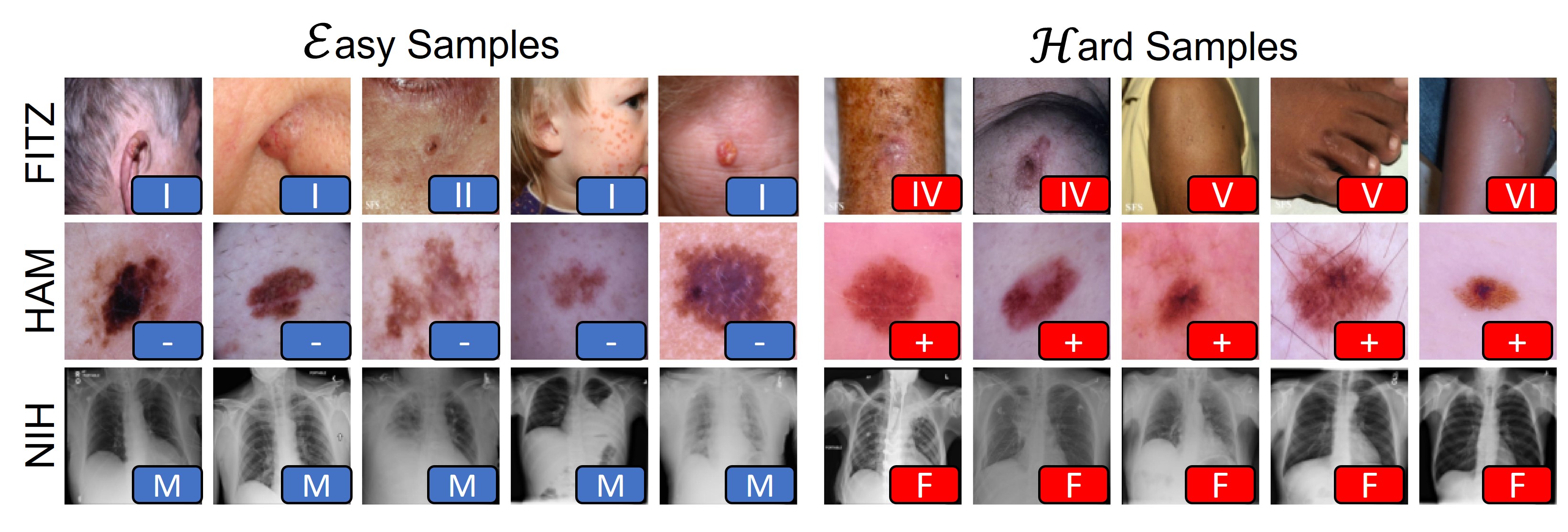}
    \caption{Visualization of easy (blue square) and hard (red square) samples across the different benchmarks. 1st row: Images from FITZ (Task 4), where each image is labeled by its Fitzpatrick skin tone, denoted as I, II, IV, V or VI. 2nd row: Images from HAM (Task 3), where each image is labeled by age (age $<$ or $\geq$ 60), denoted as $-$ or $+$, respectively. 3rd row: Images from NIH (Task 1), where each image is labeled as male or female, denoted as M or F, respectively. We notice that the hard samples represent the minority within their respective tasks.}
    \label{babc}
\end{figure}
\begin{figure}[H]
    \centering
    \includegraphics[width=1\linewidth]{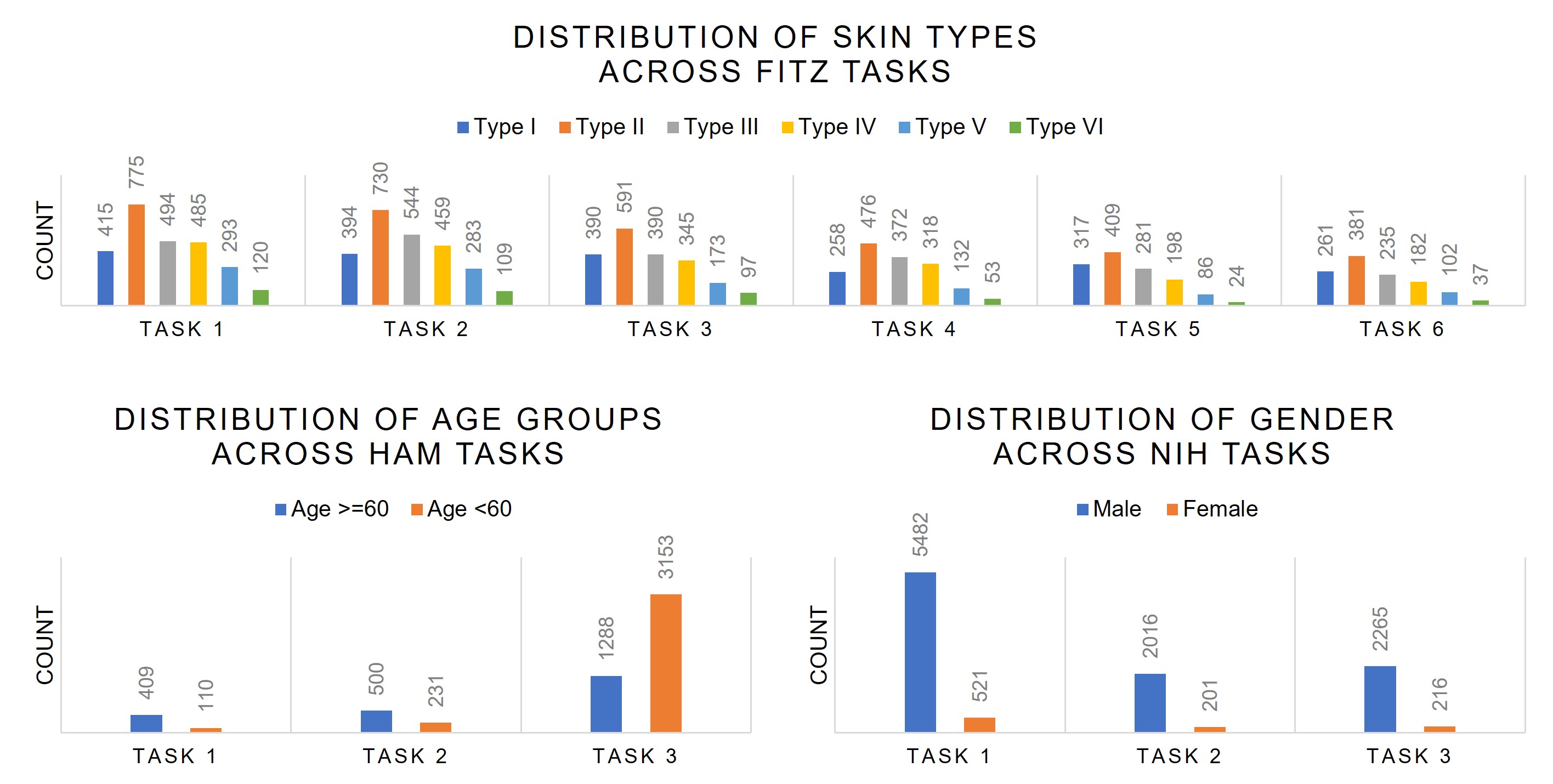}
    \caption{Bias distribution across the different tasks in FITZ, HAM and NIH.}
    \label{appendix_plot}
\end{figure}

\begin{table}[]
\centering
\caption{Distribution of images across train, validation, and test sets for each task in FITZ, HAM, and NIH. The $V$ value between  brackets next to Train represents the Cramer's $V$ correlation between the sensitive attribute (e.g., skin tone in FITZ) and disease classes. }
\resizebox{\textwidth}{!}{%
\begin{tabular}{c|cccccc|ccc|ccc}
\hline
 & \multicolumn{6}{c|}{\textbf{FITZ}} & \multicolumn{3}{c|}{\textbf{HAM}} & \multicolumn{3}{c}{\textbf{NIH}} \\
 \hline
&  Task 1 &	Task 2	&Task 3 & Task 4 &	Task 5	&Task 6	&Task 1 &	Task 2	&Task 3 & Task 1 &	Task 2	&Task 3		 \\ \hline
Train ($V$) & 
\begin{tabular}[c]{@{}c@{}}2,582 \\ (0.311)\end{tabular} & 
\begin{tabular}[c]{@{}c@{}}2,519 \\ (0.287)\end{tabular} & 
\begin{tabular}[c]{@{}c@{}}1,986 \\ (0.277)\end{tabular} & 
\begin{tabular}[c]{@{}c@{}}1,609 \\ (0.333)\end{tabular} & 
\begin{tabular}[c]{@{}c@{}}1,315 \\ (0.293)\end{tabular} & 
\begin{tabular}[c]{@{}c@{}}1,198 \\ (0.328)\end{tabular} &
\begin{tabular}[c]{@{}c@{}}519 \\ (0.187)\end{tabular} & 
\begin{tabular}[c]{@{}c@{}}731 \\ (0.112)\end{tabular} & 
\begin{tabular}[c]{@{}c@{}}4,441 \\ (0.199)\end{tabular} &
\begin{tabular}[c]{@{}c@{}}6,003 \\ (0.259)\end{tabular} & 
\begin{tabular}[c]{@{}c@{}}2,217 \\ (0.242)\end{tabular} & 
\begin{tabular}[c]{@{}c@{}}2,481 \\ (0.264)\end{tabular} \\
Val		&361&	358	&286	&230&	180&	171 & 	85	&123&	809 & 	1,711	&646	&710 \\ 
Test &	747&	721&	568&	462&	375&	344& 	167&	240&	1,563 & 3,472	&1,311&	1,442 \\ \hline
Total &	3,690&	3,598&	2,840&	2,301&	1,870	&1,713 & 771&	1,094&	6,813 & 	11,186 	&4,174 &	4,633
	\\ \hline
\end{tabular}
}
\label{nih1}
\end{table}

\begin{table}[t]
\centering
\scriptsize
\caption{Continuation of Table 2 --Standard deviation results.}
\begin{tabular}{c|c|cccccccccc}
\hline 
\multirow{2}{*}{Exp~}  & \multirow{2}{*}{Method}  & \multirow{2}{*}{F} &  \multicolumn{7}{c}{ACC}   & \multirow{2}{*}{DPR} & \multirow{2}{*}{EOD} \\ \cline{4-10}
 &  &   & Type-I &	Type-II&	Type-III&	Type-IV&	Type-V	&Type-VI	&Overall &		&  \\ \hline
 \multicolumn{12}{c}{\textbf{Comparison against Baselines}} \\ \hline
 \multirow{3}{*}{$\mathcal{A}$}  & JOINT &0.81	&	0.28	&	0.06	&	0.85	&	0.28	&	0.35	&	0.89	&	0.69	&	0.66	&	0.51   \\ 
& SINGLE & 	0.73	&	0.13	&	0.15	&	0.30	&	0.45	&	0.53	&	0.19	&	0.11	&	0.12	&	0.38
\\
& SeqFT &  2.24	&	1.68	&	2.11	&	2.48	&	2.34	&	2.58	&	2.64	&	2.33	&	2.56	&	2.84
\\ \hline
 \multicolumn{12}{c}{\textbf{Comparison against CL Methods}} \\ \hline
\multirow{3}{*}{$\mathcal{B}$} & EWC & 1.55	&	1.03	&	1.19	&	1.15	&	1.83	&	1.78	&	1.21	&	1.11	&	1.22	&	1.05   \\ 
 & PackNet & 0.21	&	0.57	&	0.41	&	0.93	&	0.25	&	0.81	&	0.35	&	0.69	&	0.76	&	0.49
 \\ 
&  SupSup & 0.98	&	0.66	&	0.52	&	0.63	&	0.55	&	0.82	&	0.54	&	0.44	&	0.38	&	0.79
 \\ \hline
\multicolumn{12}{c}{\textbf{Comparison against CL with Bias Mitigation Methods}} \\ \hline
\multirow{6}{*}{$\mathcal{C}$} & EWC+S &  1.51	&	1.76	&	1.34	&	1.19	&	1.75	&	1.26	&	1.33	&	1.28	&	1.01	&	1.14
 \\ 
& PackNet+S &  0.15	&	0.14	&	0.62	&	0.51	&	0.26	&	0.17	&	0.64	&	0.25	&	0.46	&	0.26
\\
&SupSup+S & 0.25	&	0.58	&	0.61	&	0.63	&	0.24	&	0.5	&	0.28	&	0.56	&	0.63	&	0.65
\\
&EWC+W & 1.37	&	1.27	&	1.44	&	1.68	&	1.34	&	1.25	&	1.42	&	1.62	&	2.53	&	1.48
 \\ 
& PackNet+W &  0.28	&	0.34	&	0.58	&	0.35	&	0.22	&	0.23	&	0.31	&	0.37	&	0.34	&	0.71
\\ 
&SupSup+W  &  0.63	&	0.94	&	0.41	&	0.37	&	0.82	&	0.21	&	0.52	&	0.16	&	0.35	&	0.81
  \\ \hline
\multicolumn{12}{c}{\textbf{Our Proposed Fair Continual Learning Method}} \\ \hline
$\mathcal{D}$ &\modelname{} &0.54	&	0.38	&	0.15	&	0.84	&	0.44	&	0.39	&	0.71	&	0.24	&	0.33	&	0.41
  \\\hline \hline
\multicolumn{12}{c}{\textbf{[Upper-bound] Comparison against a Bias Mitigation Method}} \\ \hline
$\mathcal{E}$& FairDisCo &   0.91	&	0.22	&	0.84	&	0.65	&	0.74	&	0.33	&	0.85	&	0.12	&	0.35	&	0.34
 \\ \hline
\end{tabular}
\label{results_fitz_std}
\end{table}

\begin{table}[t]
\centering
\scriptsize
\caption{Continuation of Table 3 --Standard deviation results.  }
\begin{tabular}{c|c|cccccc|cccccc}
\hline
\multirow{3}{*}{\rotatebox{90}{Exp}} & \multirow{3}{*}{Method}  &  \multicolumn{6}{c|}{HAM}   & \multicolumn{6}{c}{NIH} \\ \cline{3-14}
 &  & \multirow{2}{*}{F} &  \multicolumn{3}{c}{ACC}   & \multirow{2}{*}{DPR} &  \multirow{2}{*}{EOD} & \multirow{2}{*}{F} &  \multicolumn{3}{c}{ACC}   & \multirow{2}{*}{DPR} &  \multirow{2}{*}{EOD} \\ \cline{4-6} \cline{10-12}
  &  &  & ~$<$60~ &$\geq$60&	Overall &	&  &~~ &  ~~~M~~ &	F&	Overall &	&  \\ \hline
 \multicolumn{14}{c}{\textbf{Comparison against Baselines}} \\ \hline
  \multirow{3}{*}{$\mathcal{F}$} & JOINT  & 0.05	&	0.14	&	0.11	&	0.14	&	0.09	&	0.08	&	0.15	&	0.08	&	0.11	&	0.22	&	0.61	&	0.23
\\   
& SINGLE &  0.53	&	0.72	&	0.23	&	0.91	&	0.34	&	0.82	&	0.91	&	0.79	&	0.44	&	0.27	&	0.22	&	0.42
\\ 
& SeqFT &  2.64	&	0.65	&	0.85	&	0.36	&	0.41	&	0.45	&	0.54	&	0.49	&	0.96	&	0.46	&	0.22	&	0.36
\\ \hline
\multicolumn{14}{c}{\textbf{Comparison against CL Methods}} \\ \hline
\multirow{3}{*}{$\mathcal{G}$}  & EWC& 1.46	&	1.12	&	1.69	&	1.79	&	1.14	&	1.16	&	1.74	&	1.62	&	1.42	&	1.66	&	1.19	&	1.77	 \\
& PackNet& 0.44	&	0.57	&	0.28	&	0.22	&	0.82	&	0.61	&	0.22	&	0.43	&	0.51	&	0.17	&	0.34	&	0.55
	 \\
& SupSup&0.53	&	0.64	&	0.22	&	0.74	&	0.91	&	0.84	&	0.39	&	0.53	&	0.55	&	0.69	&	0.22	&	0.43
	\\ \hline
\multicolumn{14}{c}{\textbf{Comparison against CL with Bias Mitigation Methods}} \\ \hline

\multirow{6}{*}{$\mathcal{H}$}  & EWC+S & 1.64	&	1.83	&	1.72	&	1.11	&	1.96	&	1.66	&	1.36	&	1.19	&	1.57	&	1.21	&	1.44	&	1.36   \\ 
&PackNet+S  &  0.43	&	0.37	&	0.64	&	0.87	&	0.74	&	0.22	&	0.42	&	0.33	&	0.35	&	0.19	&	0.43	&	0.35

\\
&SupSup+S & 0.18	&	0.44	&	0.61	&	0.86	&	0.34	&	0.63	&	0.91	&	0.53	&	0.28	&	0.39	&	0.38	&	0.19

 \\ 
&EWC+W &  1.17	&	1.75	&	1.81	&	1.62	&	1.56	&	1.38	&	1.35	&	1.86	&	1.54	&	1.81	&	1.71	&	1.26 \\ 
 &PackNet+W & 0.84	&	0.35	&	0.71	&	0.23	&	0.55	&	0.14	&	0.45	&	0.82	&	0.22	&	0.65	&	0.27	&	0.33
 \\ 
&SupSup+W   & 0.72	&	0.43	&	0.16	&	0.44	&	0.78	&	0.81	&	0.62	&	0.77	&	0.25	&	0.34	&	0.64	&	0.81
 \\ \hline
\multicolumn{14}{c}{\textbf{Our Proposed Fair Continual Learning Method}} \\ \hline
$\mathcal{I}$ & \modelname{}  &  0.44	&	0.18	&	0.57	&	0.21	&	0.38	&	0.25	&	0.35	&	0.41	&	0.27	&	0.39	&	0.44	&	0.62
 \\ \hline \hline
\multicolumn{14}{c}{\textbf{[Upper-bound] Comparison against a Bias Mitigation Method}} \\ \hline
$\mathcal{J}$ & FairDisCo  & 0.92	&	0.64	&	0.38	&	0.39	&	0.22	&	0.37	&	0.72	&	0.21	&	0.34	&	0.61	&	0.66	&	0.74
 \\  \hline
\end{tabular}
\label{results_nih_std}
\end{table}
\end{document}